\documentclass[10pt,twocolumn,superscriptaddress,english,pre,aps,reprint,showpacs]{revtex4-1}
\usepackage{amsmath}
\usepackage{graphicx}
\usepackage{epstopdf}
\usepackage{caption}
\usepackage{subcaption}

\usepackage{babel}

\usepackage{subfig}
\makeatother

\begin{document}

\title{Kinetics of Deposition of Oriented Superdisks}

\author{B. N. Aleksi\'{c}}
\affiliation{Texas A\&M University at Qatar, Doha, Qatar}
\affiliation{Institute of Physics Belgrade, University of Belgrade, Belgrade, Serbia}
\author{N.M. \v{S}vraki\'{c}}
\affiliation{Institute of Physics Belgrade, University of Belgrade, Belgrade, Serbia}
\author{M. Beli\'{c}}
\affiliation{Texas A\&M University at Qatar, Doha, Qatar}

\begin{abstract}
We use numerical Monte Carlo simulation to study kinetics of deposition
of oriented superdisks, bounded by the Lame curves of the form $|x|^{2p}+|y|^{2p}=1$,
on regular planar substrate. It was recently shown that the maximum
packing density, as well as jamming density $\rho_{J}$, exhibit discontinuous
derivative at $p=0.5$, when the shape changes from convex to concave
form. By careful examination of the late-stage approach to the jamming
limit, we find that the leading term in temporal development is also
nonanalytic at $p=0.5$, and offer heuristic excluded-area arguments
for this behavior.
\end{abstract}


\maketitle

Deposition, or adsorption, of extended objects at different surfaces
is of considerable interest for a wide range of applications in biology,
nanotechnology, device physics, physical chemistry, and materials
science \cite{privmanNonequilibrium,privmanColloids}. Typically,
such objects range in size from submicrometer scale down to nanometers,
and, depending on the application in question, the objects could be
polymers, globular proteins, nanotubes, DNA segments, or general geometrical
shapes such as discs, polygons, etc. Early studies have focused on
deposition of simple regular shapes (lines, needles) on spatially
homogeneous, regular substrates \cite{evans} and the main issue was
the effect of shape, size, orientation, and symmetry of the depositing
objects on the late stage kinetics of this process.

Typical physical situation is that the particles (objects) are randomly
deposited on the target surface with uniform flux and that there are
no equilibrium processes active at the surface. Because of this, the
final density of the deposited objects, after which no additional
deposition is possible, is less than the maximum packing density.
This final, or jamming, density, $\rho_{J}$ has been recently studied
\cite{PRE79_042103gromenkoPrivman2009} for objects of different geometrical
shapes and symmetries, and it was shown that $\rho_{J}$ exhibits
singularity when the shape changes from convex to concave. In a separate
study of densest packing of non-overlapping objects \cite{jiaoStillingerTorquato},
the maximal packing density is found to be nonanalytic at the point
when objects become noncircular. Additionally, it was found that the
change of shape away from rotational symmetry influences packing characteristics
in nontrivial way.

Theoretically, several models have been developed to capture the basic
physics of deposition, and by far the most studied is that of random
sequential adsorption (RSA)\cite{evans}. In this model particles
(objects) are sequentially deposited on the randomly chosen site on
the substrate. When deposited, such objects are irreversibly and permanently
attached to that site. If the randomly chosen site for deposition
is already occupied, or the objects overlap due to their size or shape, the
deposition is rejected, the particle is discarded, and the deposition
is next attempted at a different randomly chosen site. Note that,
in this process, object-object and object-substrate interactions are
modeled solely by geometrical and other features included in the deposition
procedure \cite{privmanAdhes,wangNielabaPrivmanPA,araujoCadilhePrivman,araujoCadilhe,nielabaPrivman,MPLwangNielabaPrivman1993,bartleltPrivman,JPCcadilheAraujoPrivman2007,gromenkoPrivmanGlasser,PRE79_011104gromenkoPrivman2009,GonzalezHemmerHoye}.

Two main properties are of particular experimental and theoretical interest in this process: (i) the final jamming density of objects and (ii) the leading temporal approach to the jammed state. This second property prominently features in experimental situation where it is important to understand the process of approaching various specific morphological properties as the deposit is formed, for example, avoiding contact. Examples include depositions on prepatterned surfaces \cite{minkoMuller,lonovMinko,kiriyGorodyska}, or nanoparticle sintering \cite{layaniMagdassi} or in inkjet printing technology \cite{balMcMurran} where the evolution of surface morphology plays the major role. 
While the properties of the final jammed state as a function of geometry and size of the depositing objects are reasonably well understood \cite{PRE79_042103gromenkoPrivman2009}, the kinetic properties of the approach to this state have not been extensively studied (see, for instance, concluding remarks in \cite{JPCcadilheAraujoPrivman2007}). It is the purpose of present work to address this question.

\begin{figure}
\centering

\begin{minipage}{2.75cm}
\includegraphics[width=3cm,height=3cm]{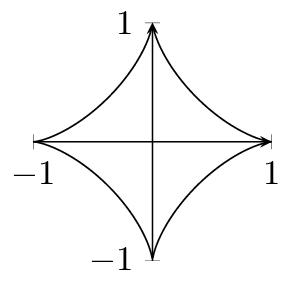}
\subcaption{$p=0.35$}
\end{minipage}
\begin{minipage}{2.75cm}
\includegraphics[width=3cm,height=3cm]{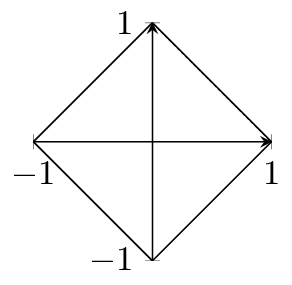}
\subcaption{$p=0.5$}
\end{minipage}
\begin{minipage}{2.75cm}
\includegraphics[width=3cm,height=3cm]{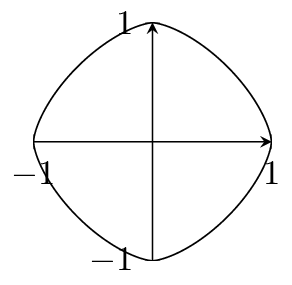}
\subcaption{$p=0.75$}
\end{minipage}

\begin{minipage}{2.75cm}
\includegraphics[width=3cm,height=3cm]{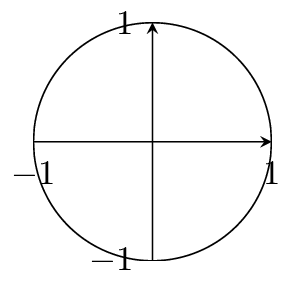}
\subcaption{$p=1.0$}
\end{minipage}
\begin{minipage}{2.75cm}
\includegraphics[width=3cm,height=3cm]{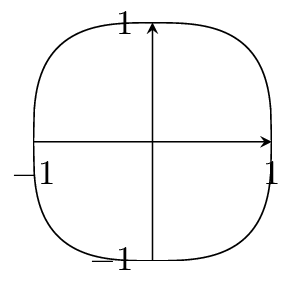}
\subcaption{$p=1.5$}
\end{minipage}
\begin{minipage}{2.75cm}
\includegraphics[width=3cm,height=3cm]{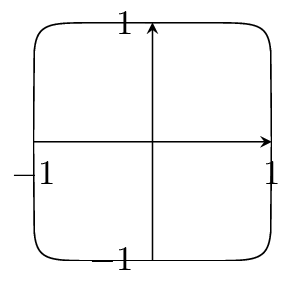}
\subcaption{$p=5.0$}
\end{minipage}

\caption{Superdisk shapes for different values of deformation parameter $p$.
Note the change of shape from concave to convex at $p=0.5$. }

\label{superdisksPic}

\end{figure}

In this work we consider the kinetics of deposition of oriented ``superdisks\textquotedblright{}
on homogeneous planar substrates. Such shapes in two dimensions are
defined by the expression $|x|^{2p}+|y|^{2p}\leq1$, where $p$ is
deformation parameter with values $p\in\left(0,\infty\right)$. As
$p$ varies the shape changes from ``cross\textquotedblright{} ($p=0$),
to square ($p=0.5$), to circular ($p=1$), to ``diamond\textquotedblright{}
($p=\infty$) as illustrated in Fig. \ref{superdisksPic}.

Note that, for $p<0.5$, the depositing object are concave, of general
astroid shape, and then become convex for $p>0.5$. Clearly, $p=0.5$
is the special point in this respect (see below). Also, when $p=1$
the shape is a circle. For all $p>1$ we get the family of objects
with quadratic symmetry, while for $p<1$ the resulting family of
objects also have quadratic symmetry, but rotated by $\pi/4$
with respect to the first group\cite{jiaoStillingerTorquato}. It
is clear, generally, that by changing the deformation parameter $p$,
one can control both convexity and the symmetry of the object.

The deposition proceeds via standard RSA algorithm: the point on the
surface is randomly chosen to place the center of the object. If the
object can ``fit\textquotedblright{} at this point without overlapping
with neighboring deposited objects it is placed there. Otherwise it
is discarded, and the next deposition is attempted at a different
randomly chosen point. In our simulation we only used oriented objects.

At the very early stages of deposition, we expect that the density
of deposited objects will grow linearly with time, since almost any
randomly chosen point for deposition attempt will be unoccupied. However,
as more and more objects are deposited, it becomes harder to find
the available spot to place an object without overlap, and the density
of deposited objects, $\rho(t)$, grows much more slowly, until it
reaches jamming density, $\rho_{J}$, after which no more objects
can be placed on the substrate. This final jammed-state configuration
is not unique and objects do not cover the substrate with maximum
packing density\cite{evans}. Indeed, the jamming density, or the
density of deposited particles per unit area, is simply related to
the jamming coverage, or the fraction of the covered area by the simple
relation, $A(p)\rho_{J}(p)$, where $A(p)$ is the area of the ``superdisk\textquotedblright{}
(see bellow), and we have explicitly indicated $p$-dependence of
the relevant quantities. Of course, the details of the approach to
the jammed state, i.e. the late-stage deposition kinetics, as well
as the jamming density itself, will depend on the shape of the objects.
Recall that this shape can be continually tuned by varying the deformation
parameter p. As mentioned, in a recent study Gromenko and Privman
\cite{PRE79_042103gromenkoPrivman2009} have shown that the jamming
density, $\rho_{J}$, is nonanalytic at $p=0.5$, i.e. when the object's
shape changes from convex to concave.

Turning back to the approach to the jamming density, recall that,
in RSA, the process is described by the standard Pomeau \cite{pomeau}
and Swendsen \cite{Swendsen} conjecture which gives asymptotic results
for oriented simple shapes that are in agreement with numerical simulations
\cite{brosilowZiffVigil}. This conjecture, however, may not be correct
for non-oriented objects \cite{privmanWangNielabaPRB,tarjusViot}. In the rest
of this paper we describe the numerical procedure used and the results
obtained for the deposition kinetics of oriented superdisks with the
deformation parameter in the range $0\leq p\leq1$.

In our Monte Carlo (MC) procedure, the substrate was of the size $500D*500D$,
where $D$ is the typical ``diameter\textquotedblright{} of the depositing
object \cite{jiaoStillingerTorquato}. In our simulation $D=2$. Once
the point for the center of the object is randomly selected on this
substrate, we tested if the whole object can ``fit\textquotedblright{}
without overlap with neighboring objects already deposited. It's not
necessary to check overlapping with all neighbors but only with nearby
points with centers not closer than $D$ from the new point. Overlap
testing is computationally ``expensive'' operation, so this approach
makes simulation much faster than checking all previously deposited
points for every new candidate point. We made two implementations,
one in ``C'' and one in ``Java'' programming language. It appears
that ``Java'' is more reliable as random number generator.

With very simple objects, like line segments and the like, this test
is straightforward , but becomes more involved in the case of superdisks.
Namely, each superdisk of the form $|x|^{2p}+|y|^{2p}\leq1$ has the
area

\begin{equation}
A(p)=4\frac{\Gamma^{2}(1+\frac{1}{2p})}{\Gamma(1+\frac{1}{p})}\label{A_p}
\end{equation}
where $\Gamma$ is standard Gamma function. No point encompassed by
this area can be shared with any other superdisk (non-overlap condition).
This means that there is an exclusion region around the deposited
superdisk within which no center of another superdisk can be placed.
For example, when $p=1$, the object is a circle with unit radius,
with area $A(1)=\pi$ , and the corresponding exclusion region is
also a circle, but with radius equal to $2$, and area $4\pi$ .

In general, the exclusion region for the convex superdisks of the
form $|x|^{2p}+|y|^{2p}\leq1$ is another superdisk of the form $|x|^{2p}+|y|^{2p}\leq2^{2p}$,
and the area of this region is $A_{cx}=4A(p)$ where $A(p)$ is given
by (\ref{A_p}), and the subscript ``cx\textquotedblright{} indicates
that $p\geq0.5$, i.e. the object is convex. Within this region, no
center of another superdisk can be placed.

\begin{figure}
\centering
\begin{minipage}{4cm}
\includegraphics[width=4cm,height=4cm]{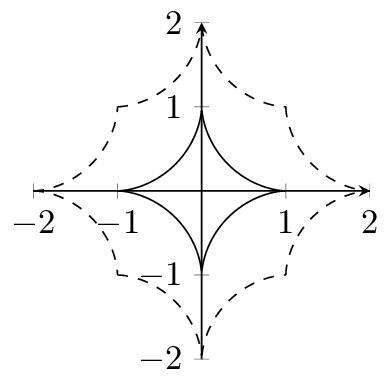}
\subcaption{}
\end{minipage}
\begin{minipage}{4cm}
\includegraphics[width=4cm,height=4cm]{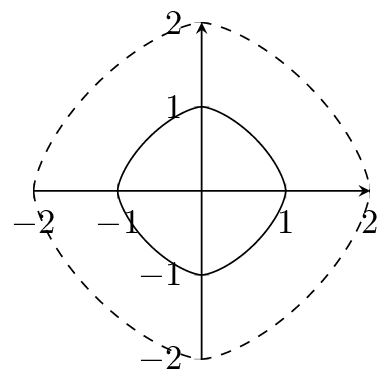}
\subcaption{}
\end{minipage}
\caption{Exclusion regions for concave and convex superdisks (outer envelopes).
The example illustrates cases with (a) p=0.3, and (b) p=0.75.}
\label{exlusionRegionPics}
\end{figure}

For concave superdisks the situation is slightly more involved. The
exclusion region is shown in Fig. \ref{exlusionRegionPics}. Its boundary
consists of the external envelope formed by four superdisks centered
at the corners of the original superdisk $|x|^{2p}+|y|^{2p}\leq1$.

\begin{figure}
\includegraphics[width=8cm,height=7cm]{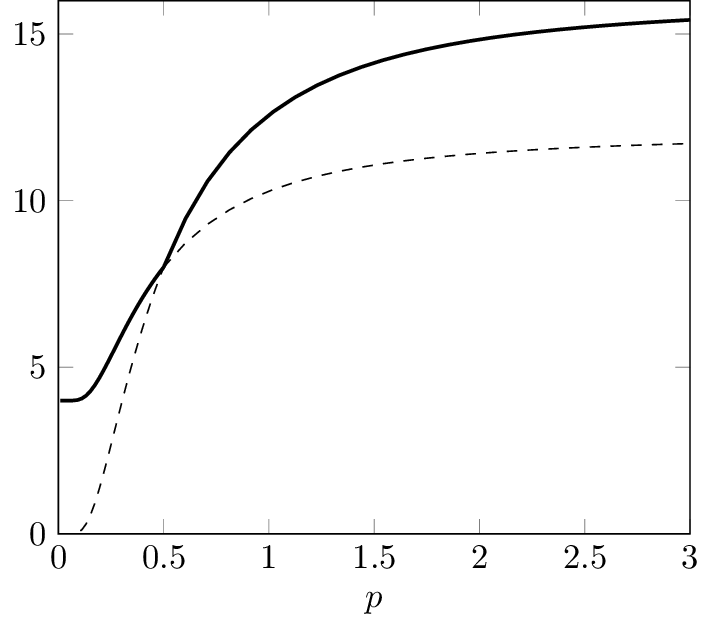}
\caption{The area (size) of the exclusion regions is marked by solid line.}
\label{areaOfER}
\end{figure}

The exclusion area of this region is easily calculated and is given
by $A_{cc}=4+2A(p)$ where $A(p)$ is given by Eq (\ref{A_p}), and
the subscript ``cc\textquotedblright{} indicates that $p\leq0.5$,
i.e. the superdisk is concave. Recall that $A(0.5)=2$, so that $A_{cx}(p=0.5)=A_{cc}(p=0.5)$,
and the exclusion areas are equal, as expected. However, the derivative
of the excluded area has discontinuity at $p=0.5$. This is clearly
visible in Fig. \ref{areaOfER}. Solid lines mark parts of $A_{cc}$ and $A_{cx}$ that must be used to determine real exclusion region. Dashed lines are their nonphysical continuations.

This nonanalyticity of the excluded area is the origin of the observed
singularity in both jammed state \cite{PRE79_042103gromenkoPrivman2009}
and the maximum packing state \cite{jiaoStillingerTorquato} of superdisks
at $p=0.5$.

To investigate the kinetics of RSA deposition of superdisks, we performed
$100$ MC runs for each value of $0\leq p\leq1$, with each run typically
of several billion steps, until the jamming limit is reached. We plausibly
expect \cite{bartleltPrivman,mannaSvrakic} that the approach of the
coverage density to its value at the jammed state is of the exponential
form:

\begin{equation}
\rho(t,p)=\rho_{J}(p)-Q(p)e^{-t\sigma(p)}\label{jammedKinetics}
\end{equation}
where $Q$ and $\sigma$ are parameters to be determined, and their
(possible) $p$-dependence is explicitly indicated. The normalized
jamming limit shows nonanalytic behavior at $p=0.5$, as explained
above \cite{PRE79_042103gromenkoPrivman2009}. The plot of normalized jamming limit faitfully reproduces the result previously obtained by Gromenko and Privman illustrated in Fig. 3. of reference \cite{PRE79_042103gromenkoPrivman2009}.

This can be also seen on the plot of the coverage density, shown below
in Fig. \ref{coverageDensityPic}.

\begin{figure}
\includegraphics[width=8cm,height=7cm]{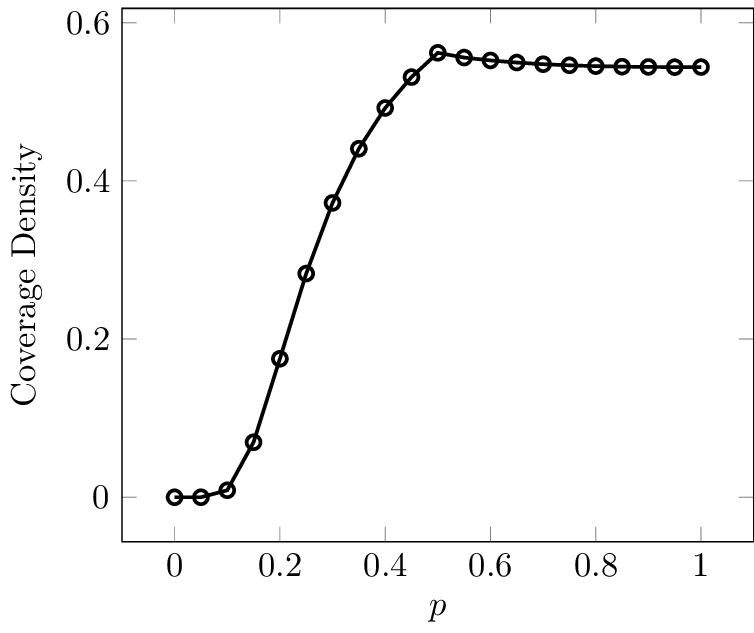}
\caption{Coverage density as a function of deformation parameter $p$.}
\label{coverageDensityPic}
\end{figure}

\begin{figure}
\includegraphics[width=8.5cm,height=7cm]{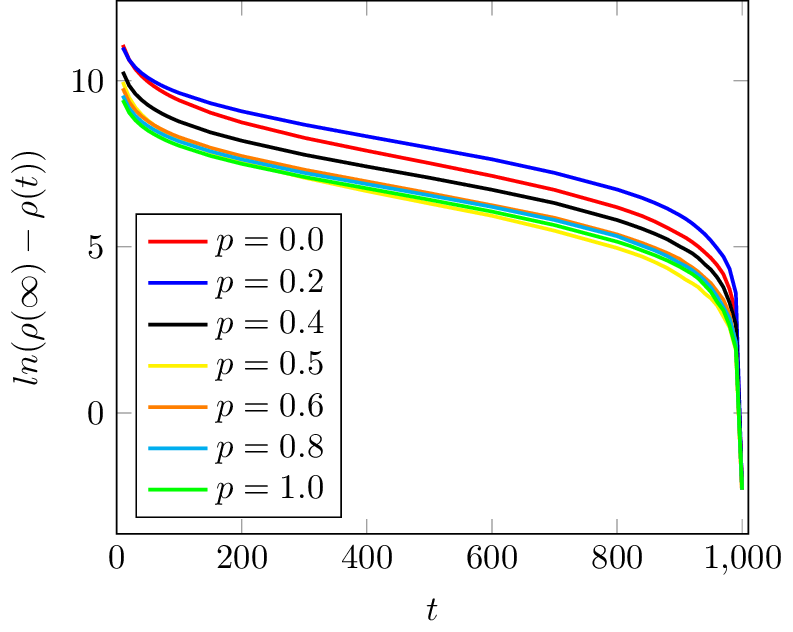}
\caption{Approach to the jamming density with time for several values of the
deformation parameter p. The linear part can be approximated by the
line of the form $-\sigma t+ln(Q(p))$}
\label{lnJammingDensityPic}
\end{figure}

Turning to the time-dependence, we plot $ln(\rho_{J}-\rho(t))$ vs.
time, for several values of parameter p, as shown on Fig. \ref{lnJammingDensityPic}.

Inspection of this graph shows: (i) that the parameter $\sigma$ from
Eq (\ref{jammedKinetics}) has no dependence on $p$, and we get $\sigma=3.5*10^{-3}$
, and (ii) that $Q(p)$, the pre-factor in front of the exponential
approach in Eq. (\ref{jammedKinetics}), depends on the deformation
parameter p in a nontrivial way.

\begin{figure}
\includegraphics[width=8cm,height=7cm]{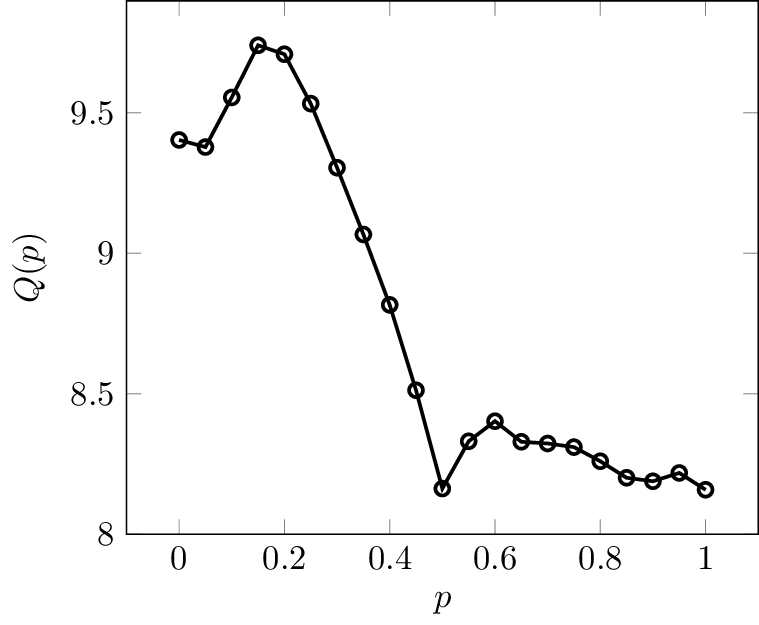}
\caption{Plot of the prefactor of the exponential approach to the jamming limit,
Eq. (\ref{jammedKinetics}), shows nontrivial dependence on $p$,
and a singularity at $p=0.5$.}
\label{prefactorPic}
\end{figure}

This is not entirely unexpected, as it is known \cite{mannaSvrakic}
that, with the RSA deposition of line segments on the substrate, the
corresponding pre-factor is inversely proportional to the length of
the segment. In this simple situation the exclusion area is simply
the length of the segment. In our case, the connection is more complicated.
The plot of $Q(p)$ vs. $p$, shown in Fig. \ref{prefactorPic}.

This term is clearly nonanalytic at $p=0.5$ (square object). We believe
that this reflects non-analyticity of the exclusion area for this
value of $p$, which is, itself, consequence of the change in convexity
of the deposited objects.

In conclusion, we have studied time dependence of deposition of oriented
superdisks of various shapes on homogeneous substrates. Our results
indicate that, in addition to maximum packing density and jamming
density, the leading term in late stage deposition also shows nonanalytic
behavior when depositing objects change their shape from convex to
concave. Intuitively, one would expect that the convexity of the depositing
object would become more important in late stage deposition kinetic,
as they pack closer and closer, and the details of their shape and
the size of the exclusion area begins to play more prominent role.
The study of this effect for non-oriented superdisks is currently
under way. 
\begin{acknowledgments}
This publication was made possible by NPRP grants \# 09 - 462 - 1 - 074 and \# 5 - 674 - 1 -114 from the Qatar National Research Fund (a member of Qatar Foundation). The statements made herein are solely the responsibility of the authors. Work
at the Institute of Physics Belgrade is supported by the Ministry
of Science of the Republic of Serbia under the projects OI 171006
and ON 171017.\end{acknowledgments}

\end{document}